\documentclass[prb,twocolumn,showpacs,preprintnumbers,amsmath,amssymb]{revtex4}
\usepackage{epsf}
\usepackage{graphicx}
\usepackage{dcolumn}
\usepackage{bm}
\usepackage{epstopdf}

\begin{document}

\title{Resonant 2D-2D tunneling with account for spin-orbit interaction}

\author{I.V. Rozhansky}
\email{rozhansky@gmail.com}
\affiliation{Ioffe Institute, 194021 St. Petersburg, Russia}
\affiliation{Lappeenranta University of Technology, P.O. Box 20,
FI-53851, Lappeenranta, Finland}
\author{N.S. Averkiev}
\affiliation{Ioffe Institute, 194021 St. Petersburg, Russia}
\author{E. L\"ahderanta}
\affiliation{Lappeenranta University of Technology, P.O. Box 20,
FI-53851, Lappeenranta, Finland}
\date{\today}

\begin{abstract}
We present a theory of quantum tunneling between 2D layers
with account for Rashba and Dresselhaus spin-orbit interaction (SOI) in the layers.
Energy and momentum conservation results in a single resonant peak in the tunnel conductance between two 2D layers as has been experimentally observed for two quantum wells (QW) in GaAs/AlGaAs
heterostructures. The account for SOI in the layers leads
to a complex pattern in the tunneling characteristic with typical features corresponding to SOI energy. For this manifestation of SOI to be observed experimentally the characteristic energy should exceed
the resonant broadening related to the particles quantum lifetime in the layers.
We perform an accurate analysis of the known experimental data on electron and hole 2D-2D tunneling
in AlGaAs/GaAs heterostructures. It appears that for the electron tunneling
the manifestation of SOI is difficult to observe, but for the holes tunneling
the parameters of the real structures used in the experiments are very close to
those required by the resolution criteria.
We also consider a new promising candidate for the effect to be observed, that is p-doped SiGe strained heterostructures. The reported parameters of cubic Rashba SOI and quantum lifetime in strained Ge QWs fabricated up to date
already match the criteria for observing SOI in 2D-2D heavy holes tunneling.
As supported by our calculations small adjustments of the parameters for AlGaAs/GaAs p-type QWs
or simply designing a 2D-2D tunneling experiment for SiGe case are very likely to reveal the SOI features in the 2D-2D tunneling.
\end{abstract}


\maketitle

\index{Rozhansky I. V.}
\index{Averkiev N. S.}
\section{Introduction}

Quantum tunnelling between two 2D layers separated by a weakly transparent potential barrier is a prime example of resonant phenomena in semiconductor nanostructures.
Energy and in-plane momentum conservation put tight restrictions on the 2D-2D tunneling so that
the conductance between identical layers exhibits delta function-like maximum at zero bias, while at any finite voltage applied across the barrier the
quantum transitions
between the layers are forbidden. The phenomena has been observed experimentally for both n-doped and p-doped
AlGaAs/GaAs heterostructures with two spatially separated QWs \cite{Eisenstein1995,Eisenstein2007}.
The resonant peak appearing in the tunneling I-V characteristic at zero bias is
broadened to the characteristic width $\Gamma$ due to scattering in the layers, for QW
of a high quality the electron-electron scattering is probably the main contribution \cite{MacDonald1993}.
With account for spin-orbit interaction (SOI) in the 2D layers the tunneling transport turns out to be more complex.
SOI splits the size quantization subbands in the layers allowing the resonant tunneling between one spin-orbit subband of the left layer into
another subband of the right layer. This can occur when the left and right states has the same energy, i.e. when the spin-orbit splitting is exactly compensated by an external bias. It was demonstrated theoretically that in this way SOI would manifest itself in the tunnel conductance with
two resonant peaks or more complex pattern with typical features shifted from zero bias by a characteristic
energy of the spin-orbit splitting \cite{Raichev2003,ZyuzinRaikh2006}.
For exactly the same 2D layers the tunneling between opposite spin-orbit subbands (assuming no spin-flip during the tunneling)
would be forbidden because of the orthogonality of the spin wavefunctions. However,
in real heterostructures with two QWs two doping layers are located below the lowest QW and above the highest,
so that the ionized dopant layers create electrical fields
which have opposite directions in the two QWs.
Consequently, the parameter of Rashba SOI which is proportional to the electric field has the opposite sign in the two QWs so that (in the absence of other spin-orbit contributions) the spin-orbit subbands in one layer are reversed compared to the other one, the spin structure of the upper subband in one layer matches   that of the lower subband in the other one. This removes the restriction on the
tunneling between the opposite subbands. Introducing SOI of Dresselhaus type (the same in both layers) results in a more complex spin structure of the eigenstates in the layers and more rich tunneling pattern as a result of interference between the two SOI contributions.
In our previous works we considered electron tunneling between two n-type 2D layers (n-n tunneling) with account for both Rashba and Dreseelhaus contributions and
obtained analytical expression for the tunneling current in this general case \cite{Rozhansky2008,Rozhansky2009}. However, the question why the SOI effects had not been seen in the tunneling experiments remained open.
For the manisfestation of SOI in the 2D-2D tunneling transport to be observed the SOI characteristic energy $\delta$ should exceed the resonance broadening $\Gamma$.
In this paper we generalize the theory and consider the tunneling of heavy holes between two p-type layers (h-h tunneling). We consider parameters of the heterostructures used in the experiments on n-n and h-h tunneling and analyze the requirements for the SOI pattern to be resolved experimentally.
Our calculations show that for the p-type AlGaAs/GaAs heterostructures studied experimentally the parameters are very close to the SOI features to be observed in the tunnel current.
Moreover, for the p-type case there is a strong scaling with doping level and QW thickness so that a small correction of the parameters must lead ro the experimental observation of SOI in the tunneling transport.
Another promising candidate for the  2D-2D tunneling experiment
is the SiGe strained heterostructures. The strained Ge QWs possess only Rashba SOI, the Dresselhaus term is absent due to lack of bulk inversion asymmetry for Ge lattice. We consider two different heterostructures with p-type QWs studied experimentally\cite{Moriya2014,Myronov2015} for which an accurate
data on the SOI interaction was obtained. The SOI in these structures is larger than in AlGaAs case so designing a tunneling experiment is very attractive as our calculations confirm.
\begin{figure}[h]
\includegraphics[width=0.4\textwidth]{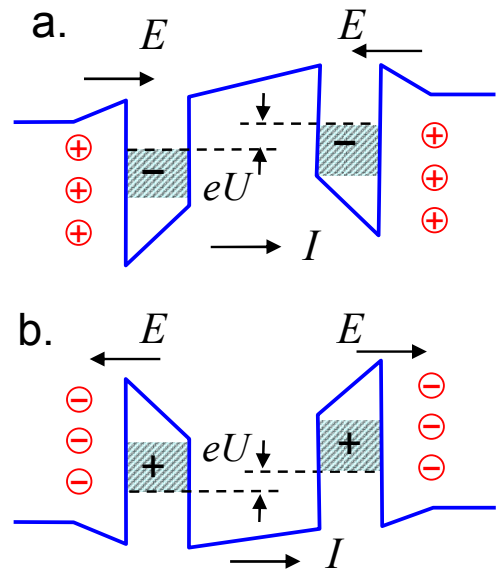}
\caption{(Color online) Energy diagram of two 2D layers.
The n-n tunneling case (a) is realized by donor impurity doping layers located as illustrated by red
circles. The electrons (dashed regions) populate QWs formed by conduction band profile. An external bias $U$  is applied to the QWs which results in the tunneling current $I$. The electric field $E$ formed by the ionized donor layers has opposite direction in the two QWs.
The h-h tunneling case (b) occurs when the doping of p-type. In this case
the current between the QWs formed by the valence band potential profile is carried by holes.
The electric field in this case $E$ is formed by the ionized acceptor layers with negative charge.
}
\label{fig:layers}
\end{figure}

\section{Theory}
The system under consideration is illustrated by Fig.~\ref{fig:layers}.
We consider two QWs separated by a weakly transparent barrier. Keeping in mind the relevant experiments \cite{Eisenstein1995,Eisenstein2007} the QWs could be of GaAs grown along [001] direction (z-axis), in-plane axis are $x\parallel\text{[100]}$, $y\parallel\text{[010]}$, the barriers formed by  Al$_{x}$Ga$_{1-x}$As.
Depending on the doping the QWs can be popualted either by electrons or by holes. The first case is shown
in Fig.~\ref{fig:layers}(a). Here the QWs for the electons are formed by the conductance band profile.
The doping is made by the two layers of donors as illustrated by red circles.
The ionized donors create a positive charge and
an electric field directed as shown in the figure.
The difference between the chemical potentials of the electrons in the QWs is created by an external bias $U$ applied directly to the QWs which
enables the tunneling of the elctrons from the right QW to the left QW with electric current flowing in the opposite direction.
For the case of h-h tunneling the situation is quite similar as shown in Fig.~\ref{fig:layers}(b).
The figure shows the valence band profile and
the ionized acceptors form
two negatively charged layers so that the direction of the electric field is opposite to that of electron case, the current flows in the same direction as the holes flux.
We consider one size quantization level for the electrons for n-n tunneling and one size quantization level for the heavy holes (HH1) in the case of h-h tunneling, zero temperature is assumed.
We use the Bardeen's tunneling Hamiltonian approach \cite{MacDonald1993} with the total Hamiltonian of the system expressed as:
\begin{equation}
\label{HT0} H=H^L+H^R+H_T,
\end{equation}
where $H^L,H^R$ are the partial Hamiltonians for the left
and right layers respectively, $H_T$ is the tunneling term.
Assuming spin is conserved during the tunneling the current density is given by\cite{Rozhansky2009}:
\begin{equation}
\label{eqKubo}
j = \frac{{eT^2
}}{{4\pi^3\hbar ^3 }}{\mathop{\rm Re}\nolimits} \left\{ \mathrm{Tr}
\int { {G_{}^R \left( \bf{p},\varepsilon  - eU \right)G_{}^L
\left( \bf{p},\varepsilon  \right)d{\bf{p}}d\varepsilon } }.
\right\},
\end{equation} where $\varepsilon$ is the electron energy, $\bf{p}$ is its in-plane momentum, $G^{R,L}$ are Green's functions of the left and
right layers respectfully, $T$ is the Bardeen's tunneling matrix element, the measure of the tunnel
coupling between the layers, $e$ is the elementary charge. With electron
spin taken into account the Green's functions are 2x2 metrices.
The spinor basis states are defined by angular momentum projection on z-axis, for the electrons
the basis is $(1/2,-1/2)$ while for the heavy holes it is $(3/2,-3/2)$.
The Green's functions for the electrons or holes in the layers are to be calculated with account for SOI and scattering processes in the layers.
As long as the interlayer transitions rate is decribed by the formula (\ref{eqKubo})
the Green's functions in (\ref{eqKubo}) are aware of intralayer interactions only, any
inter-layer scattering processes which could give rise to vertex corrections shouldn't be taken into account in the leading order of the tunneling parameter\cite{Rozhansky2009}.
We consider two types of SOI Hamiltonians for the electron and holes tunneling repsectively.
For the n-n tunneling both Rashba and Dresselhaus contributions to SOI in a 2D layer are linear in the in-plane momentum:
\begin{equation}
\label{eqHSOe}
 H^{SO}_e  = \alpha_e \left( {k_y
\sigma _x  - k_x \sigma _y } \right) + \beta_e \left( {k_x \sigma _x -
k_y \sigma _y } \right)
\end{equation}
where $k_x,k_y$ are the in-plane wavevector components, $\sigma_x,\sigma_y$ are Pauli matrices.
z-axis are normal to the QWs planes, $\alpha_e$ and $\beta_e$ are the parameters of
Rashba and Dresselhaus contributions to SOI. The Rashba term is assumed to be linear in
external electric field $\alpha_e\sim E$.
For the h-h tunneling between the considered QWs
the dominating contribution to the Rashba SOI is cubic in the in-plane wavevector\cite{WinklerBook}, while
the Dresselhaus part is linear.
We use the following Hamiltonian for the valence band Rashba\cite{Moriya2014} and Dresselhaus\cite{DurnevGlazov2014} terms:
\begin{equation}
\label{eqHSOh}
 H^{SO}_h  = \alpha_h i \left( {k_ - ^3 \sigma _ +   - k_ + ^3 \sigma _ -  } \right) + \beta_h \left( {\sigma _x k_x  + \sigma _y k_y } \right),
\end{equation}
where $\sigma_\pm=\left(\sigma_x\pm i \sigma_y\right)/2$, $k_{\pm}=k_x\pm i k_y$.
With no other interactions taken into account the Hamiltonian for a 2D layer takes the form
\begin{equation}
\hat H  =  - \frac{{\hbar ^2 }}{{2m }}\left( {\frac{{d^2 }}{{dx^2 }} + \frac{{d^2 }}{{dy^2 }}} \right) + \hat H^{SO},
\end{equation}
where $m$ is the in-plane mass for the relevant carriers.
Its eigenvalues $\varepsilon$ and eigenvectors $\phi$ give the two spin-orbit subbands:
\begin{equation}
\label{eqEigen}
\varepsilon_\mp  = \frac{{\hbar ^2 k^2 }}{{2m}} \pm \left| u \right|,\,\,\,\
\phi _ \mp   = \frac{1}{{\sqrt 2 }}\left( {\begin{array}{*{20}c}
   \pm 1  \\
   {\gamma ^* }  \\
\end{array}} \right),
\end{equation}
where for the electrons
\[
u = u_e \equiv \alpha_e ik_ -   + \beta_e k_+,
\]
for the heavy holes
\[
u = u_h \equiv \alpha_h ik_ - ^3  + \beta_h k_- ,
\]
for both cases $\gamma={u}/{\left|u\right|}$.
We will now assume that intralayer scattering processes do not involve spin so that
the only spin-dependent intercation is the SOI described by (\ref{eqHSOe}) or (\ref{eqHSOh}).
Therefore, in the basis of the eigenstates $(\phi_- ,\phi_+)$ the single-particle Green's function
of the carrier in a 2D layer is expressed as:
\[
G  = \left[ {\begin{array}{*{20}c}
   {G_ -  } & 0  \\
   0 & {G_ +  }  \\
\end{array}} \right],
\]
where
\begin{equation}
G_\pm\left(k,\varepsilon\right)={\frac{1}{{\varepsilon  +\varepsilon_F-
\frac{{\hbar ^2 k^2 }}{{2m}} \pm |u|   + i\frac{\hbar }{{2\tau
}}\text{sgn}\;\varepsilon }}},
\end{equation}
$\varepsilon_F$ being the Fermi level of the layer in the absence of applied voltage.
The quantum lifetime $\tau$ incorporates all the scattering processes in the layer.
It was suggested that in the experiments on 2D-2D resonant tunneling in AlGaAs based heterostructures \cite{Eisenstein1995,Eisenstein2007} the key contribution to a particle quantum lifetime $\tau$ comes
from the e-e or h-h Coulomb scattering.
In the standard spinor basis the same Green's function matrix is:
\begin{equation}
\label{eqGmain}
G = \frac{1}{2}\left[ {\begin{array}{*{20}c}
   {G_ -   + G_ +  } & {\gamma \left( {G_ -   - G_ +  } \right)}  \\
   {\gamma ^* \left( {G_ -   - G_ +  } \right)} & {G_ -   + G_ +  }  \\
\end{array}} \right].
\end{equation}
Inserting (\ref{eqGmain}) into (\ref{eqKubo}) and assuming the spin-orbit splitting small compared ro
the Fermi energy we end up with the expression for the tunneling current density:
\begin{align}
\label{eqCurrentMain}
&j = \frac{{e mT^2 }}{{4\pi ^2 \hbar ^3 }}\times\nonumber\\
&\int\limits_0^{2\pi }  \left[ \begin{aligned}
& \left( {\frac{eU\Gamma}{{\left( {eU - \xi _ -  } \right)^2  + \Gamma ^2 }} + \frac{eU\Gamma}{{\left( {eU + \xi _ -  } \right)^2  + \Gamma ^2 }}} \right)\left( {1 + {\mathop{\rm Re}\nolimits} \gamma ^L \gamma ^{R*} } \right) \\
 & + \left( {\frac{eU\Gamma}{{\left( {eU - \xi _ +  } \right)^2  + \Gamma ^2 }} + \frac{eU\Gamma}{{\left( {eU + \xi _ +  } \right)^2  + \Gamma ^2 }}} \right)\left( {1 - {\mathop{\rm Re}\nolimits} \gamma ^L \gamma ^{R*} } \right)
 \end{aligned} \right]{d\varphi }
\end{align}
where
\begin{align}
& \Gamma=\frac{\hbar}{\tau} \nonumber \\
& \gamma ^{L,R}  = \frac{{u^{L,R} }}{{\left| {u^{L,R} } \right|}} \nonumber\\
& \xi _ \pm   = \left| {u^R \left( {k_F ,\varphi } \right)} \right| \pm \left| {u^L \left( {k_F ,\varphi } \right)} \right| ,
 \end{align}
indices $L$ and $R$ denote left and right layer, respectfully, $\varphi$ is the polar angle for the in-plane wavevector ${\mathbf k}=(k,\varphi)$, $k_F=\sqrt{2m\varepsilon_F}/\hbar$.
In the case of no SOI or if it is identical in the two layers, i.e.
$u^L=u^R$ (\ref{eqCurrentMain}) reduces to the well-known result \cite{MacDonald1993}:
\begin{equation}
\label{eqj0}
j = \frac{{2emT^2 \Gamma }}{{\pi \hbar ^3 }}\frac{{eU}}{{\left( {eU} \right)^2  + \Gamma ^2 }}.
\end{equation}
Another limitting case appears when we take Rashba terms in the two layers being of the same magnitude but of the opposite sign and completely neglect Dresselhaus terms \cite{ZyuzinRaikh2006,Rozhansky2008}.
That is
\begin{equation}
\label{eqjRashba}
j = \frac{{emT^2 \Gamma }}{{\pi \hbar ^3 }}\left( {\frac{{eU}}{{\left( {eU - 2\delta } \right)^2  + \Gamma ^2 }} + \frac{{eU}}{{\left( {eU + 2\delta } \right)^2  + \Gamma ^2 }}} \right),
\end{equation}
where
\begin{align*}
\label{eqenRashba}
& \delta  = \alpha _e k_F\,\,\,\text{(for electrons)}\\
& \delta  = \alpha _h k_F^3\,\,\,  \text{(for holes)}.
\end{align*}
Instead of a single resonance at zero bias in the abscence or identical SOI (\ref{eqj0}) now we have two resonances located at $2\delta$.
Adding a finite Dresselhaus contribution (equal in both layers) makes the picture more
complex as a result of interference between the spin structure of left and right layers states.
Typical resonant features in the tunneling I-V characteristic still occur at a bias corresponding to Rashba energy $\delta_R$ given by (\ref{eqenRashba}) and Dresselhaus chracaterisric energy, $\delta_D  =\beta_{e,h}k_F$.
The so-called 'spin helix'
 case\cite{Salis2012} corresponds to equaly strong Rashba and Dresselhaus contributions,
 that is
\begin{align*}
& \beta _e  = \left| {\alpha _e } \right|      \,\,\,&\text{(for electrons)}\\
& \beta _h  = \left| {\alpha _h } \right|k_F^2 \,\,\, &\text{(for holes)}
 \end{align*}
The current density in this case is given by:
\begin{align}
&j = \frac{{emT^2 eU}}{{\pi \hbar ^3 }}\left( {\frac{{\left| {{\mathop{\rm Im}\nolimits} \sqrt {g_ -  } } \right|}}{{\left| {g_ -  } \right|}} + \frac{{\left| {{\mathop{\rm Im}\nolimits} \sqrt {g_ +  } } \right|}}{{\left| {g_ +  } \right|}}} \right) \nonumber\\
&g_ \pm   = \left( {eU - i\Gamma } \right)^2  \pm 8\delta ^2 \nonumber\\
&\delta=2\beta_{e,h}k_F.
\end{align}
The resonant features in this case appear at $eU=\pm2\sqrt{2}\delta$ with rather unusial
behaivour, a decrease in $\Gamma$ suppresses the current \cite{Rozhansky2008}.

From experimental point of view the complex resonant pattern in the I-V curve would be resolved only if the characteristic SOI energy $\delta$ exceeds the broadening $\Gamma$.
As will be shown below this is not the case for the existing experiments on n-n\cite{Eisenstein1995} and h-h\cite{Eisenstein2007} tunneling which explains why the pattern related to SOI was not revealed. However, for the h-h tunneling in particular, only slight adjustment of the structure parameters would immediately lead to a well resolved fine structure of I-V characteristic related to SOI.


\section{Estimates and calculations}
\subsection{AlGaAs/GaAs heterostructures}
\begin{figure}[h]
\includegraphics[width=0.4\textwidth]{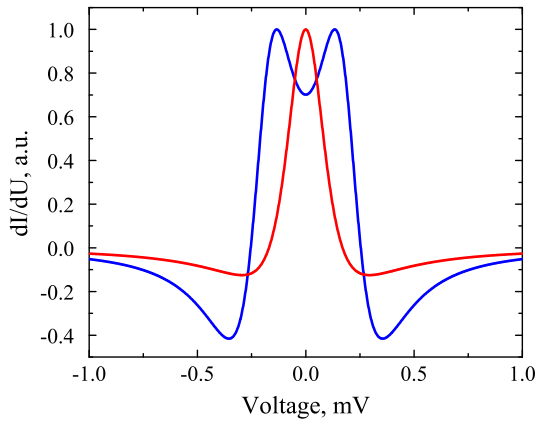}
\caption{(Color online) The calculated tunneling differential conductivity for n-n tunneling case. Red curve represents calculation with the parameters relevant to the experiment\cite{Eisenstein1995} listed in Table \ref{table1}, blue curve corresponds to the sheet electron density being increased up to $n\sim 10^{12}$ cm$^{-2}$.
}
\label{Fig2}
\end{figure}

Let us review the relevant parameters affecting the position of SOI related resonant peaks on I-V curve and the broadening in a real experimental situation. Everywhere below we assume Rashba parameters having opposite sign in the two layers due to the opposite direction of the electric field as illustrated by Fig.~\ref{fig:layers}, that is $\alpha_{e,h}^L=-\alpha_{e,h}^R$. Dresselhaus SOI parameters are considered the same for both layers $\beta_{e,h}^L=\beta_{e,h}^R$.
For the n-n tunneling we rely on the SOI parameters
reported for AlGaAs/GaAs heterostructures with a QW of 12 nm thickness and
n-type Si doping in the barrier\cite{Salis2012} as these structures are rather similar to those used in the n-n tunneling experiment\cite{Eisenstein1995}.
 The Rashba parameter and the sheet electron density in these samples are listed below:
\begin{align}
\label{eqexperSalis}
&\alpha_e\approx2\cdot10^{ - 11} \,\text{eV}\cdot\text{cm} \nonumber\\
&n \approx 5\cdot10^{11} \,\text{cm}^{ - 2}
\end{align}
The electric field created by the charged plane of the ionized donors layer (Fig.~\ref{fig:layers}(a)) can be estimated as (SI units):
\begin{equation}
\label{eqEQW}
E = \frac{en}{{2\varepsilon \varepsilon _0 }},
\end{equation}
where $\varepsilon$ is the dielectric constant of the QW material, $e$ is the elementary charge, $\varepsilon_0$ is the vacuum permittivity.
With $\varepsilon=12$ for GaAs at low temperature from (\ref{eqexperSalis}) and (\ref{eqEQW}) we get
$E \approx 4\cdot10^4$ V/cm.
The Rashba coefficient $\alpha_e$ is proportional to the electric field $\alpha_e=r_eE$, from the values above we
get:
\begin{equation}
r_e\approx5\cdot10^{-16} \,e\cdot\text{cm}^2.
\end{equation}
We further apply this value to the experimental data on 2D n-n tunneling \cite{Eisenstein1995}. Knowing the sheet electron density we obtain the value for Rashba parameter and Rashba spin-orbit energy $\delta_R=\alpha_e k_F$.
For an estimate of Dresselhaus parameter $\beta_e$ we use the value obtained in the same experiment
\cite{Salis2012} as is and neglect its obviously weaker dependence on the QW thickness, the corresponding Dresselhaus characteristic energy is $\delta_D=\beta_e k_F$.
The values used for the analysis of the n-b tunneling experiment are summarized in Table \ref{table1} together with the quantum lifetime measured as the resonance broadening $\Gamma=\hbar/\tau$ in Ref. \cite{Eisenstein1995}.
  \begin{table}
  \begin{tabular}{ l | c | c }
      \hline
 Parameter & {Electrons (n-n)} & {Holes (h-h)} \\ \hline
Sheet density  &$ 1.6\cdot{10^{11}}\,\text{cm}^{-2}$&$ 7.2\cdot{10^{10}}\,\text{cm}^{-2}$\\
\hline
Rashba parameter  $\alpha_{e,h}$ &${\rm{6}} \cdot {\rm{1}}{{\rm{0}}^{ - 12}}\,{\rm{ eV}} \cdot {\rm{cm}}$ & $1.3 \cdot {10^{ - 22}}\,{\rm{eV}} \cdot {\rm{c}}{{\rm{m}}^{\rm{3}}}$ \\ \hline
Dresselhaus parameter  $\beta_{e,h}$ &${\rm{3}} \cdot {\rm{1}}{{\rm{0}}^{ - 11}}\,{\rm{ eV}} \cdot {\rm{cm}}$ & $3.6 \cdot {10^{ - 10}}\,{\rm{eV}} \cdot {\rm{c}}{{\rm{m}}}$ \\ \hline

Rashba energy $\delta_R$ & $0.005$ meV & $0.05$ meV\\ \hline
Dresselhaus energy $\delta_D$ & $0.03$ meV & $0.25$ meV\\ \hline
Broadening $\Gamma=\hbar/\tau$ &$0.17$ meV & $0.17$ meV\\ \hline
  \end{tabular}
  \caption{List of parameters relevant to the experiments on n-n tunneling \cite{Eisenstein1995} and h-h tunneling \cite{Eisenstein2007} between two QWs in AlGaAs/GaAs based heterostructure.}
  \label{table1}
    \end{table}
As can be seen from Table \ref{table1}, for the n-n tunneling experiment\cite{Eisenstein1995} the
resonance broadening $\Gamma$ substantially exceeds the spin-orbit energies $\delta_R,\delta_D$, so
 that the SOI features could not be observed. Fig.~\ref{Fig2} shows the normalized differential conductivity
 calculated using (\ref{eqCurrentMain}) for the n-n tunneling with the parameter set listed in Table \ref{table1} (red curve). There's no visible difference from the case of the tunneling with no SOI at all.
Let us now address the question how the parameters are to be changed for the SOI to be revealed.
Note that the Rashba energy $\delta_R$ is proportional to the electric field and Fermi wavevector $k_F$,
so it scales with the sheet electron density as $
\delta_R\sim n^{3/2}$, the scaling for Dresselhaus SOI is $\delta_D\sim n^{1/2}$. We will assume $\Gamma$ independent of $n$ (if we considered the electron-electron scattering to dominate, its scaling would be\cite{Giuliani1982} $\Gamma\sim 1/n$, however other scattering processes could start playing a role with sheet density varied).  Thus to make the SOI larger than $\Gamma$ the sheet electron density must be increased by almost an order of magnitude from $n\sim 10^{11}$ cm$^{-2}$ to
$n\sim 10^{12}$ cm$^{-2}$. Blue curve in Fig.~\ref{Fig2} shows the calculation for this particular case.
The SOI-related resonances are rather well-resolved. Whether the required substantial increase of the doping level can be realized technologically without affecting other structure parameters (including the dominating of e-e scattering) is not that clear
but the task does not seem completely unrealistic.
As we will show below the situation for the holes is far more favorable.

\begin{figure}[h]
\includegraphics[width=0.4\textwidth]{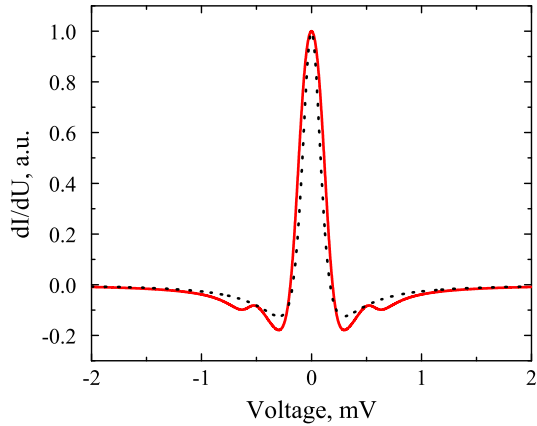}
\caption{(Color online) The calculated tunneling differential conductivity for h-h tunneling case. Red curve represents calculation with the parameters relevant to the experiment\cite{Eisenstein2007} listed in Table \ref{table1}, the dotted curve shows the case of no SOI in the layers.
}
\label{Fig3}
\end{figure}

\begin{figure}[h]
\includegraphics[width=0.4\textwidth]{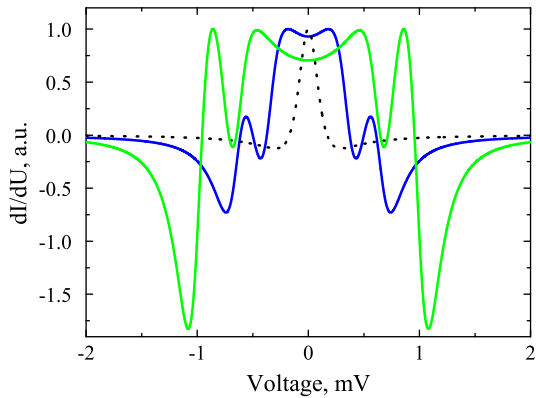}
\caption{(Color online) The calculated tunneling differential conductivity for h-h tunneling case with sheet hole density increased by a factor of 2 (green curve) up to $p=1.4\cdot 10^{11}\,\text{cm}^{-2}$ and for the QW thickness increased up to $a=20$ nm (blue curve), the dotted curve shows the case of no SOI in the layers.
}
\label{Fig4}
\end{figure}

For the case of h-h tunneling we take an estimate for $r_h=\alpha_h/E$ from Ref.\cite{Winkler2002} for 20 nm thick QW. That is
\begin{equation}
\label{eqrh20}
r_{h}(20\text{ nm})  \approx 7.5\,\cdot10^{ - 26} \,e\cdot{\rm{cm}}^4
\end{equation}
Unlike electron case the Rashba SOI for the heavy holes depends on the separation between
LH and HH subbands in the QW, therefore it is supposed to be strongly dependent on the QW thickness $a$.
Assuming infinitely deep QW this dependence is $r_{h}\sim 1/a^4$ (valid while the perturbation theory holds, i.e. the SO splitting energy is smaller than the LH-HH level separations)\cite{Winkler2002,WinklerBook}. Thus, for a narrower
15 nm thick QW used in h-h tunneling experiments we obtain:
\begin{equation}
\label{eqRh}
r_{h}(15\text{ nm})=r_{h}(20\text{ nm})\left(\frac{15}{20}\right)^4\approx 2.4\,\cdot10^{ - 26} \,e\cdot{\rm{cm}}^4
\end{equation}
An estimate for Dresselhaus SOI contribution for the HH1 subband in a GaAs QW grown along [001] appears to be less reliable as there is no clear experimental data here.
From the theoretical calculations with account for QW interface mixing contribution the Dresselhaus
parameter for the holes HH1 level in a 10 nm thick GaAs [001] QW can be estimated as\cite{DurnevGlazov2014}
$\beta_h\approx3.6\cdot10^{-10}\,\text{eV}\cdot\text{cm}$.
With the sheet hole density reported for the experiment\cite{Eisenstein2007} $p=7.2\cdot10^{10}\,\text{cm}^{-2}$ and the corresponding Fermi wavevector $
k_F\approx7.4\cdot10^5 \,\text{cm}^{ - 1}$ we get the values summarized in Table \ref{table1} for the h-h tunneling experiment\cite{Eisenstein2007}.
Note that the spin-orbit energies $\delta_R,\delta_D$ are by an order of magnitude higher
than those for n-n tunneling case and they are of the same order as $\Gamma$.
Fig. \ref{Fig3} shows the calculated differential conductivity for the h-h tunneling with the parameters listed in Table \ref{table1} (solid curve).
Dotted curve in Figs. \ref{Fig3},\ref{Fig4} shows the calculated conductivity with no account for SOI as a reference.
Thus, the characteristic shown in Fig. \ref{Fig3} is supposed to reproduce the experimental result reported in Ref.\cite{Eisenstein2007}.
The SOI features are almost not resolved, although a small oscillating feature at $U\approx\pm 0.3$ mV is surprisingly similar to the one seen in the experimental curve at $-0.3$ mV bias in \cite{Eisenstein2007}.
Anyway, unlike the electron case the set of parameters is very close to that required for experimental observation of SOI in the h-h tunneling.
Moreover, because the main contribution to Rashba SOI for the heavy holes is cubic in the wavevector,
the scaling on the sheet hole density $p$ is stronger than for the electrons.
With the same arguments as above we have now $\delta_R\sim p^{5/2}$. Increasing $p$ by a factor of 2 would make both $\delta_R,\delta_D$ far larger than $\Gamma$ so that all the details including the interference  between Rashba and Dresselhaus contributions would be clearly resolved.  This is demonstrated by calculation result shown in Fig.~\ref{Fig4}.
Also due to the strong dependence of $\alpha_h$ on the QW thickness a similar enhancement in $\delta_R$ would be achieved if the QW width thickness was increased from $15$ nm up to $20$ nm.
The Dresselhaus parameter $\beta_h$ is also known to depend on the QW thickness\cite{DurnevGlazov2014}, however not that strongly as the Rashba term, we neglect this dependence in the calculations for the 20 nm thick QW also shown in Fig.~\ref{Fig4}.
Note that the patterns in Fig.~\ref{Fig4} are somewhat different. Changing the QW thickness does not affect  the Dresselhaus energy $\delta_D$ while changing the sheet density affects $\delta_D$ through $k_F$.
\subsection{Ge/SiGe heterostructures}
\begin{figure}[h]
\includegraphics[width=0.4\textwidth]{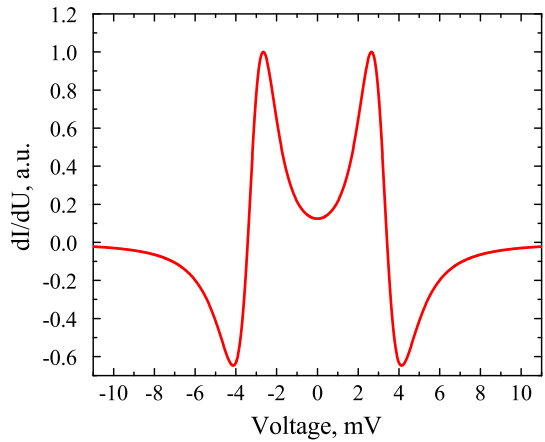}
\caption{(Color online) The calculated tunneling differential conductivity for h-h tunneling case using the parameters of the strained Ge QW studied in \cite{Myronov2015} listed in Table \ref{table2} (Experimnet 2).
}
\label{Fig5}
\end{figure}

Another promising material for the SOI to be observed in the 2D-2D tunneling is a SiGe based heterostructure.
 Having a central-symmetric lattice Ge semiconductor do not have Dresselhaus SOI, thus for Ge QWs
the only contribution to SOI is the Rashba term. Similarly to the AlGaAs/GaAs case discussed above SOI for the heavy holes is larger than for the electrons making
h-h tunneling experiment more attractive.
Recently strained p-type Ge QWs were fabricated having a large HH-LH subbands separation which results in a strong domination of cubic-Rashba SOI\cite{Moriya2014,Myronov2015}.
In the experiment\cite{Moriya2014} the SOI splitting $\delta_h=\alpha_hk_F^3$ was reported to be $0.3-0.4$ meV. The corresponding value of
$r_h=2.3\cdot10^{-27}\,e\cdot\text{cm}^4$, is an order of magnitude smaller than for GaAs QW (\ref{eqRh}) which is the direct consequence of the
LH-HH separation being larger than in GaAs QW.
However, due to higher sheet density and additional external electric field Rashba energy in these structures is nearly an order of magnitude higher than that listed in Table \ref{table1} for h-h tunneling in GaAs QWs.
Unfortunately, the gain in $\delta_h$ appears to be less than the loss in the broadening $\Gamma$. The transport time in the QWs studied in Ref.\cite{Moriya2014} was $\tau_{tr}\sim 0.2$ ps which gives a lower bound for $\Gamma>\hbar/\tau_{tr}\sim 3$ meV. At that $\delta_h/\Gamma$ is well below 1 so the SOI-related structure in 2D-2D tunneling cannot be resolved. An example of a typical set of parameters from Ref.\cite{Moriya2014} is listed in Table \ref{table2} (Experiment 1).
\begin{table}
  \begin{tabular}{ l | c | c }
      \hline
 Parameter & Experiment 1 & Experiment 2\\ \hline
Rashba parameter  $\alpha_h$ &${\rm{1.6}} \cdot {\rm{10}}^{ - 23}\,{\rm{ eV}} \cdot {\rm{cm}}$ & $5 \cdot {10^{ - 23}}\,{\rm{eV}} \cdot {\rm{c}}{{\rm{m}}^{\rm{3}}}$ \\ \hline
Rashba energy $\delta_R$ & $0.35$ meV & $1.6$ meV\\ \hline
Broadening $\Gamma=\hbar/\tau$ &$>3.7$ meV & $1.3$ meV\\ \hline
  \end{tabular}
  \caption{Rashba parameters and resonance broadening for strained Ge QWs studied in Ref.\cite{Moriya2014} (Experiment 1) and in Ref.\cite{Myronov2015} (Experiment 2).
QW thickness is $a\approx 11$ nm in both cases.}
  \label{table2}
    \end{table}
On the contrary, the Ge QWs studied in Ref.\cite{Myronov2015} are already suitable for the tunneling experiment.
The difference from \cite{Moriya2014} is in smaller LH-HH separation (for the same QW thickness) due to optimized strain which leads to a strong enhancement of $\alpha_h$.
Fig.~\ref{Fig5} presents the calculated tunnel characteristic using the parameters of 'SiGe1' sample from Ref.\cite{Myronov2015} listed in Table \ref{table2} (Experiment 2). As there is no Dresselhaus term the current density follows the reduced expression (\ref{eqjRashba}).
As seen from the figure the two resonances at $\pm 2\delta_К$ are clearly resolved, thus a 2D-2D tunneling experiment using the two strained Ge QWs
is very likely to reveal this SOI pattern.

\section{Summary}
The tunneling between 2D layers with SOI separated by a potential barrier has a nontrivial resonant character that reflects the spin structure of eigenstates in the layers.
Consequently, the dependence of tunneling conductance on the
voltage applied across the barrier appears to be very sensitive to the parameters of SOI.  At the same time the homogeneous broadeneing of the resonant peaks due to a finite quantum lifetime of the particles in the layers smears out the interference pattern. This trade-off between the characteristic spin-orbit splitting energy and the inverse quantum lifetime is crucial for experiment. We have carefully examined
the parameters of the reported experiments on 2D-2D resonant tunneling in AlGaAs/GaAs heterostructures and calculated the tunnel current using the developed theory. The calculation confirms that the SOI interference pattern is unlikely to be resolved with the given parameters set. However, a feasible adjustment of the sheet density or the QW thickness would change the situation radically. Of particular interest is the case of the tunneling between heavy holes subbands. The SOI energy in this case is by an order of magnitude higher than for the n-n tunneling. So as we have shown increasing the sheet hole density in the QWs by a factor of 2 or increasing the QWs thickness from 15 nm to 20 nm would immediately lead to a well-pronounced fine structure of the I-V characteristic related to SOI in the QWs.
Another very promising candidate for the suggested 2D-2D tunneling experiment is the Ge strained QWs with exclusively Rashba SOI.
We considered parameters of two different heterostructures studied experimentally. The calculation of 2D-2D tunneling using the parameters of the QWs experimentally studied in Ref.\cite{Myronov2015} shows well-resolved fine structure, which is likely to be observed in the experiment. We believe that such a tunneling experiment definitely is worth considering.

\section*{ACKNOWLEDGEMENTS}
We sincerely thank J.P. Eisenstein for very fruitful discussions, 
we would also like to express our gratitude to P. Altmann. R. Moriya, M. Myronov, amd R. Winkler for appreciated discussions and comments.
\bibliography{tunneling_2d}

\begin{thebibliography}{14}
\expandafter\ifx\csname natexlab\endcsname\relax\def\natexlab#1{#1}\fi
\expandafter\ifx\csname bibnamefont\endcsname\relax
  \def\bibnamefont#1{#1}\fi
\expandafter\ifx\csname bibfnamefont\endcsname\relax
  \def\bibfnamefont#1{#1}\fi
\expandafter\ifx\csname citenamefont\endcsname\relax
  \def\citenamefont#1{#1}\fi
\expandafter\ifx\csname url\endcsname\relax
  \def\url#1{\texttt{#1}}\fi
\expandafter\ifx\csname urlprefix\endcsname\relax\def\urlprefix{URL }\fi
\providecommand{\bibinfo}[2]{#2}
\providecommand{\eprint}[2][]{\url{#2}}

\bibitem[{\citenamefont{Murphy et~al.}(1995)\citenamefont{Murphy, Eisenstein,
  Pfeiffer, and West}}]{Eisenstein1995}
\bibinfo{author}{\bibfnamefont{S.~Q.} \bibnamefont{Murphy}},
  \bibinfo{author}{\bibfnamefont{J.~P.} \bibnamefont{Eisenstein}},
  \bibinfo{author}{\bibfnamefont{L.~N.} \bibnamefont{Pfeiffer}},
  \bibnamefont{and} \bibinfo{author}{\bibfnamefont{K.~W.} \bibnamefont{West}},
  \bibinfo{journal}{Phys. Rev. B} \textbf{\bibinfo{volume}{52}},
  \bibinfo{pages}{14825} (\bibinfo{year}{1995}).

\bibitem[{\citenamefont{Eisenstein et~al.}(2007)\citenamefont{Eisenstein,
  Syphers, Pfeiffer, and West}}]{Eisenstein2007}
\bibinfo{author}{\bibfnamefont{J.}~\bibnamefont{Eisenstein}},
  \bibinfo{author}{\bibfnamefont{D.}~\bibnamefont{Syphers}},
  \bibinfo{author}{\bibfnamefont{L.}~\bibnamefont{Pfeiffer}}, \bibnamefont{and}
  \bibinfo{author}{\bibfnamefont{K.}~\bibnamefont{West}},
  \bibinfo{journal}{Solid State Communications} \textbf{\bibinfo{volume}{143}},
  \bibinfo{pages}{365 } (\bibinfo{year}{2007}).

\bibitem[{\citenamefont{Zheng and MacDonald}(1993)}]{MacDonald1993}
\bibinfo{author}{\bibfnamefont{L.}~\bibnamefont{Zheng}} \bibnamefont{and}
  \bibinfo{author}{\bibfnamefont{A.~H.} \bibnamefont{MacDonald}},
  \bibinfo{journal}{Phys. Rev. B} \textbf{\bibinfo{volume}{47}},
  \bibinfo{pages}{10619} (\bibinfo{year}{1993}).

\bibitem[{\citenamefont{Raichev and Debray}(2003)}]{Raichev2003}
\bibinfo{author}{\bibfnamefont{O.~E.} \bibnamefont{Raichev}} \bibnamefont{and}
  \bibinfo{author}{\bibfnamefont{P.}~\bibnamefont{Debray}},
  \bibinfo{journal}{Phys. Rev. B} \textbf{\bibinfo{volume}{67}},
  \bibinfo{pages}{155304} (\bibinfo{year}{2003}).

\bibitem[{\citenamefont{Zyuzin et~al.}(2006)\citenamefont{Zyuzin, Mishchenko,
  and Raikh}}]{ZyuzinRaikh2006}
\bibinfo{author}{\bibfnamefont{V.~A.} \bibnamefont{Zyuzin}},
  \bibinfo{author}{\bibfnamefont{E.~G.} \bibnamefont{Mishchenko}},
  \bibnamefont{and} \bibinfo{author}{\bibfnamefont{M.~E.} \bibnamefont{Raikh}},
  \bibinfo{journal}{Phys. Rev. B} \textbf{\bibinfo{volume}{74}},
  \bibinfo{pages}{205322} (\bibinfo{year}{2006}).

\bibitem[{\citenamefont{Rozhansky and Averkiev}(2008)}]{Rozhansky2008}
\bibinfo{author}{\bibfnamefont{I.~V.} \bibnamefont{Rozhansky}}
  \bibnamefont{and} \bibinfo{author}{\bibfnamefont{N.~S.}
  \bibnamefont{Averkiev}}, \bibinfo{journal}{Phys. Rev. B}
  \textbf{\bibinfo{volume}{77}}, \bibinfo{pages}{115309}
  (\bibinfo{year}{2008}).

\bibitem[{\citenamefont{Rozhansky and Averkiev}(2009)}]{Rozhansky2009}
\bibinfo{author}{\bibfnamefont{I.~V.} \bibnamefont{Rozhansky}}
  \bibnamefont{and} \bibinfo{author}{\bibfnamefont{N.~S.}
  \bibnamefont{Averkiev}}, \bibinfo{journal}{Low Temperature Physics}
  \textbf{\bibinfo{volume}{35}} (\bibinfo{year}{2009}).

\bibitem[{\citenamefont{Moriya et~al.}(2014)\citenamefont{Moriya, Sawano,
  Hoshi, Masubuchi, Shiraki, Wild, Neumann, Abstreiter, Bougeard, Koga
  et~al.}}]{Moriya2014}
\bibinfo{author}{\bibfnamefont{R.}~\bibnamefont{Moriya}},
  \bibinfo{author}{\bibfnamefont{K.}~\bibnamefont{Sawano}},
  \bibinfo{author}{\bibfnamefont{Y.}~\bibnamefont{Hoshi}},
  \bibinfo{author}{\bibfnamefont{S.}~\bibnamefont{Masubuchi}},
  \bibinfo{author}{\bibfnamefont{Y.}~\bibnamefont{Shiraki}},
  \bibinfo{author}{\bibfnamefont{A.}~\bibnamefont{Wild}},
  \bibinfo{author}{\bibfnamefont{C.}~\bibnamefont{Neumann}},
  \bibinfo{author}{\bibfnamefont{G.}~\bibnamefont{Abstreiter}},
  \bibinfo{author}{\bibfnamefont{D.}~\bibnamefont{Bougeard}},
  \bibinfo{author}{\bibfnamefont{T.}~\bibnamefont{Koga}}, \bibnamefont{et~al.},
  \bibinfo{journal}{Phys. Rev. Lett.} \textbf{\bibinfo{volume}{113}},
  \bibinfo{pages}{086601} (\bibinfo{year}{2014}).

\bibitem[{\citenamefont{Failla et~al.}(2015)\citenamefont{Failla, Myronov,
  Morrison, Leadley, and Lloyd-Hughes}}]{Myronov2015}
\bibinfo{author}{\bibfnamefont{M.}~\bibnamefont{Failla}},
  \bibinfo{author}{\bibfnamefont{M.}~\bibnamefont{Myronov}},
  \bibinfo{author}{\bibfnamefont{C.}~\bibnamefont{Morrison}},
  \bibinfo{author}{\bibfnamefont{D.~R.} \bibnamefont{Leadley}},
  \bibnamefont{and}
  \bibinfo{author}{\bibfnamefont{J.}~\bibnamefont{Lloyd-Hughes}},
  \bibinfo{journal}{Phys. Rev. B} \textbf{\bibinfo{volume}{92}},
  \bibinfo{pages}{045303} (\bibinfo{year}{2015}).

\bibitem[{\citenamefont{Winkler}(2003)}]{WinklerBook}
\bibinfo{author}{\bibfnamefont{R.}~\bibnamefont{Winkler}},
  \emph{\bibinfo{title}{Spin-Orbit Coupling Effects in Two-Dimensional Electron
  and Hole Systems}} (\bibinfo{publisher}{Springer Berlin Heidelberg},
  \bibinfo{year}{2003}).

\bibitem[{\citenamefont{Durnev et~al.}(2014)\citenamefont{Durnev, Glazov, and
  Ivchenko}}]{DurnevGlazov2014}
\bibinfo{author}{\bibfnamefont{M.~V.} \bibnamefont{Durnev}},
  \bibinfo{author}{\bibfnamefont{M.~M.} \bibnamefont{Glazov}},
  \bibnamefont{and} \bibinfo{author}{\bibfnamefont{E.~L.}
  \bibnamefont{Ivchenko}}, \bibinfo{journal}{Phys. Rev. B}
  \textbf{\bibinfo{volume}{89}}, \bibinfo{pages}{075430}
  (\bibinfo{year}{2014}).

\bibitem[{\citenamefont{Walser et~al.}(2012)\citenamefont{Walser, Reichl,
  Wegscheider, and Salis}}]{Salis2012}
\bibinfo{author}{\bibfnamefont{M.~P.} \bibnamefont{Walser}},
  \bibinfo{author}{\bibfnamefont{C.}~\bibnamefont{Reichl}},
  \bibinfo{author}{\bibfnamefont{W.}~\bibnamefont{Wegscheider}},
  \bibnamefont{and} \bibinfo{author}{\bibfnamefont{G.}~\bibnamefont{Salis}},
  \bibinfo{journal}{Nat. Phys.} \textbf{\bibinfo{volume}{8}},
  \bibinfo{pages}{757} (\bibinfo{year}{2012}).

\bibitem[{\citenamefont{Giuliani and Quinn}(1982)}]{Giuliani1982}
\bibinfo{author}{\bibfnamefont{G.~F.} \bibnamefont{Giuliani}} \bibnamefont{and}
  \bibinfo{author}{\bibfnamefont{J.~J.} \bibnamefont{Quinn}},
  \bibinfo{journal}{Phys. Rev. B} \textbf{\bibinfo{volume}{26}},
  \bibinfo{pages}{4421} (\bibinfo{year}{1982}).

\bibitem[{\citenamefont{Winkler et~al.}(2002)\citenamefont{Winkler, Noh, Tutuc,
  and Shayegan}}]{Winkler2002}
\bibinfo{author}{\bibfnamefont{R.}~\bibnamefont{Winkler}},
  \bibinfo{author}{\bibfnamefont{H.}~\bibnamefont{Noh}},
  \bibinfo{author}{\bibfnamefont{E.}~\bibnamefont{Tutuc}}, \bibnamefont{and}
  \bibinfo{author}{\bibfnamefont{M.}~\bibnamefont{Shayegan}},
  \bibinfo{journal}{Phys. Rev. B} \textbf{\bibinfo{volume}{65}},
  \bibinfo{pages}{155303} (\bibinfo{year}{2002}).

\end{thebibliography}
\end{document}